\documentclass{article}
\usepackage{ijcai24}

\usepackage{times}
\usepackage{soul}
\usepackage{url}
\usepackage[hidelinks]{hyperref}
\usepackage[utf8]{inputenc}
\usepackage[small]{caption}
\usepackage{graphicx}
\usepackage{amsmath}
\usepackage{amsthm}
\usepackage{booktabs}
\usepackage{algorithm}
\usepackage[switch]{lineno}

\usepackage{longtable}
\usepackage{musicography}
\usepackage{multirow}
\usepackage{algpseudocode}
\usepackage{amssymb}
\usepackage{tablefootnote}

\title{Loop Copilot: Conducting AI Ensembles for Music Generation and Iterative Editing}

\author{
Yixiao Zhang$^1$
\and
Akira Maezawa$^2$\and
Gus Xia$^{3}$\and
Kazuhiko Yamamoto$^2$\And
Simon Dixon$^1$
\\
\affiliations
$^1$C4DM, Queen Mary University of London\\
$^2$Yamaha Corporation\\
$^3$Music X Lab, MBZUAI
}

\begin{document}

\maketitle

\begin{abstract}
Creating music is an iterative process, requiring varied methods at each stage. However, existing AI music systems fall short in orchestrating multiple subsystems for diverse needs.
To address this gap, we introduce Loop Copilot, a novel system that enables users to generate and iteratively refine music through an interactive, multi-round dialogue interface. The system uses a large language model to interpret user intentions and select appropriate AI models for task execution. Each backend model is specialized for a specific task, and their outputs are aggregated to meet the user's requirements. To ensure musical coherence, essential attributes are maintained in a shared data structure.  
We evaluate the effectiveness of the proposed system through semi-structured interviews and questionnaires, highlighting its utility not only in facilitating music creation but also its potential for broader applications. \footnote{This work was done when Yixiao Zhang was an intern at Yamaha Corporation.}\footnote{Code available at \url{https://github.com/ldzhangyx/loop-copilot}.}
\end{abstract}

\section{Introduction}

Music creation is an art that has traditionally been the domain of expert human musicians. Recently, with the advent of artificial intelligence (AI) music models~\cite{survey}, the music creation process is becoming more democratized. However, in the real world, there are two significant challenges in the human music creation process: first, music creation involves multiple phased tasks, from harmony and melody crafting, to arrangement and mixing; second, music creation is an inherently iterative process that cannot be achieved in one step. It usually undergoes multiple refinements before reaching its final form. Most current AI models, including interactive music interfaces and dedicated generative models, fall short in at least one of these two challenges.

\begin{figure}[tbp]
\centering
  \includegraphics[width=\linewidth]{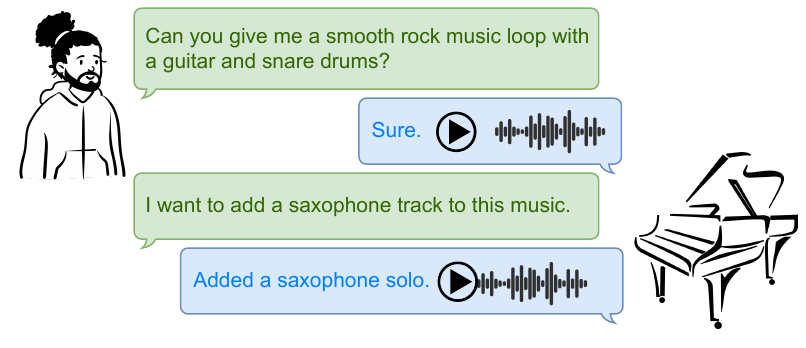}
  \caption{A conceptual illustration of interaction with Loop Copilot. The diagram depicts a two-round conversation: initially, a user requests music generation and the AI provides a loop. In the subsequent round, the user seeks modifications, and the AI offers a refined loop, emphasizing Loop Copilot's iterative feedback-driven music creation process.}
  \label{fig:teaser}
\end{figure}

Interactive music interfaces excel in melody inpainting but often lack adaptability for diverse music creation. Current popular interactive music interfaces~\cite{cococo,magentastudio,visualization} are powerful and user-friendly, but they predominantly focus on a singular type of musical modification: melody inpainting—filling in gaps based on an existing melody. These models, with their intuitive human-in-the-loop interactions, have undoubtedly lowered the entry barrier for users. However, these AI-based interfaces for music creation, although recognizing the importance of iterative generation and refinement, often rely on a single task throughout the process. This reliance not only hampers their flexibility but also restricts their adaptability to diverse music creation needs. 

On the other hand, dedicated music models offer broad capabilities but tend to have a narrow focus, limiting their application. Existing dedicated music generative models have demonstrated significant capabilities across a myriad of tasks in music creation, such as controlled music generation using chord progressions~\cite{polyffusion}, text prompts~\cite{musicgen,musiclm}, and perception~\cite{fadernets}. They also span a spectrum of music style transfer tasks at the score~\cite{polydis,accomontage}, performance~\cite{mididdsp}, and timbre~\cite{timbretransfer} levels. However, a prevalent issue with these models is their `one-off' design approach. They often treat music generation as a singular process, either focusing strictly on music generation or specific editing tasks, like style transfer. As a result, users looking to engage in a comprehensive music creation process find themselves scouting for various models to cater to different aspects of their musical needs.

In this paper, we introduce Loop Copilot, a system designed to address these challenges. It allows users to generate a music loop and iteratively refine it through a multi-round dialogue with the system. By leveraging a large language model~\cite{llmsurvey}, Loop Copilot seamlessly integrates various specialized models catering to different phases of music creation. It harnesses the power of individual models to provide a rich set of generation and editing tools. The intuitive and unified interaction is facilitated through a conversational interface, reminiscent of the benefits of the first category, while applying the strengths of the second. 

Loop Copilot is built on three key components: a large language model (LLM) controller, which interprets user intentions, selects suitable AI models for task execution, and gathers the outputs of these models; a set of backend AI models, which carry out specific tasks; and a Global Attribute Table (GAT), which records necessary music attribute information to ensure continuity throughout the creation process. Intuitively, users can utilise the LLM to `conduct' the AI ensemble, guiding the music creation process through conversation. 

In summary, our contributions 
are:

\begin{enumerate}
    \item We introduce Loop Copilot, a novel system that integrates LLMs with specialized AI music models. This enables a conversational interface for collaborative human-AI creation of music loops. 
    \item We develop the Global Attribute Table that serves as a dynamic state recorder for the music loop under construction, thereby ensuring that the musical attributes remain consistent in the iterative editing process.
    \item We conduct an interview-based comprehensive evaluation, which not only measures the performance of our system but also sheds light on the advantages and limitations of using an LLM-driven iterative editing interface in music co-creation.
\end{enumerate}

\section{Related Work}
\subsection{Music generation techniques}

Music generation has become a central topic in Music Information Retrieval (MIR) research~\cite{survey}. Both symbolic~\cite{accomontage,symbolicdiffusion,popmusictransformer} and audio-based methods~\cite{jukebox,musicldm} have been at the forefront of these efforts. As researchers ventured deeper, the desire for more control over the generation process grew~\cite{aisongcontest}. This aspiration led to the birth of controlled music generation techniques. Techniques spanned various aspects: from music structure~\cite{structure} and perception~\cite{fadernets}, to lyrics~\cite{songmass} and latent representations~\cite{ec2vae,polydis}. Particularly noteworthy are text-to-music models like Music\-LM~\cite{musiclm} and AudioLDM 2~\cite{audioldm2}, which harness text as a high-level control, marking a significant advance in user-guided music generation.

Concurrently, while the generation domain flourished, automatic music editing was emerging as a nascent, yet crucial, field. Prior works have ventured into style transfer~\cite{groove2groove,polydis}, inpainting~\cite{inpainting}, and automatic arrangement~\cite{multitrack,accomontage,accomontage2}. Recent innovations have expanded the scope to tasks like audio track addition~\cite{audit} and domain-specific instructions~\cite{instructme}. Our work aims to coordinate various tools to provide a comprehensive and flexible suite for music creation.

\subsection{Interactive music creation interfaces}

Interactive music creation interfaces have emerged as a promising avenue for harnessing the potential of AI in music creation. Some of these interfaces are built upon AI models~\cite{cococo,visualization,calliope,mysong}, while others are extensions of traditional music software~\footnote{e.g.\ Band-in-a-Box, \url{https://www.pgmusic.com/}}. CoCoCo is an improved interactive interface based on the CoCoNet model~\cite{coconet} trained using Bach's works and designed to assist users in composing music for four voices. Rau at al.~\cite{visualization} designed a new front-end interface interaction for the MelodyRNN model, where the system provides the user with multiple candidates to choose from and allows for editing at different levels of granularity. These AI-based interfaces provide varying degrees of control over the music creation process. However, their functionality is typically tied to a single backend model, which limits their adaptability and restricts the range of tasks they can support.

More similarly, COSMIC~\cite{cosmic} is a conversational system for music co-creation that leveraged several backend models, including CoCon~\cite{cocon} and BUTTER~\cite{butter} for lyrics generation and melody generation, respectively. COSMIC represented a significant step forward in interactive music creation, but it was limitated in its capabilities. Building upon the foundational ideas of COSMIC, our work integrates a Large Language Model (LLM) and broadens the range of backend models, aiming to offer a more natural and diverse user interaction experience, thereby pushing the envelope of usability.

\subsection{Large language models in music creation}

Large language models (LLMs) have found application in music creation, such as synthesizing text descriptions for music audio~\cite{lpmusiccaps} and lyrics writing~\cite{chatgptlyrics}. The advent of LLMs has opened up new possibilities for their use in music creation. LLMs have the potential to understand complex user inputs and guide multiple AI tools accordingly, enabling a more dynamic and flexible approach to music creation.

The use of LLMs as controllers to direct multiple AI tools is relatively novel. The potential of LLMs in this role has been demonstrated in a few recent studies~\cite{visualchatgpt,hugginggpt,audiogpt}, which served as the inspiration for our work. Visual ChatGPT~\cite{visualchatgpt}, as the first work to make a similar attempt, collected a number of visual models as back-end models and called them using ChatGPT; HuggingGPT~\cite{hugginggpt} further leveraged the unified API of the Huggingface Community to be able to select the appropriate model from hundreds of existing models for different tasks. Following the previous research, the LLM in our system also acts as an interpreter of user intentions, selecting suitable AI music models for task execution and integration of their outputs. This not only makes the system more intuitive and user-friendly, but also allows users to express their creative ideas more freely and directly.

In summary, our work builds upon the foundations laid by previous research in music generation, interactive music creation interfaces, and the use of LLMs in music creation. Loop Copilot brings together these elements to support human-AI co-creation in music. The novelty of our work lies in the use of an LLM to `conduct' an ensemble of AI models, thereby providing a versatile, intuitive, and user-friendly interface for iterative music creation and editing.

A parallel work is MusicAgent~\cite{musicagent}. Similar to ours, MusicAgent utilises a large language model to ensemble multiple music understanding and generation tasks. However, it falls short in interative editing, which limits its application in music production.

\section{System design}

\begin{figure*}[tbp]
    \centering
\includegraphics[width=\linewidth]{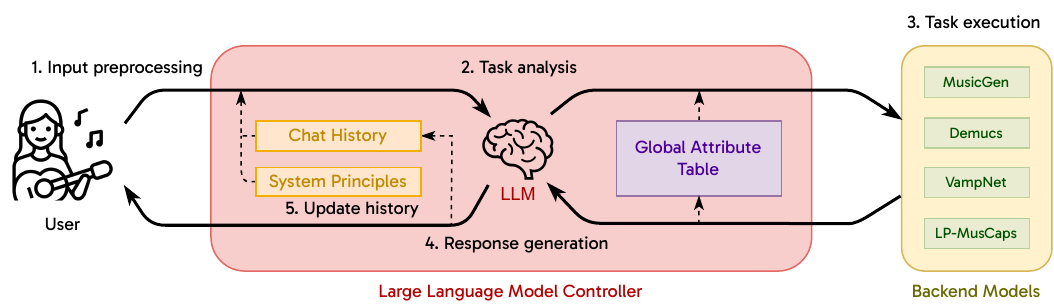}
    \caption{The diagram of Loop Copilot's workflow. Once the user inputs the request, firstly, Loop Copilot \textbf{preprocesses the input} and converts it to textual modality; secondly, the LLM, based on the input, the system principles, and the chat history, performs the \textbf{task analysis} and calls the corresponding models; after that, the backend models \textbf{execute the task} and output the result; finally, the LLM does the \textbf{final processing} of the output and returns it.}
    \label{fig:diagram}
\end{figure*}

\subsection{Model Formulation}

We begin with an example of a typical interaction process, shown in Figure \ref{fig:teaser}.  It comprises two key steps: the user initially 1) \textbf{drafts a music loop} (\textit{``Can you give me a smooth rock music loop with a guitar and snare drums?"}); and then  2) \textbf{iteratively refines it} through multiple rounds of dialogue (\textit{``I want to add a saxophone track to this music."}). After completing the 2-round dialogue, the current status can be represented by a sequence $[(Q_1, A_1), (Q_2, A_2)]$, where $Q$ means the user's question and $A$ means the answer of Loop Copilot.

To formally define the interaction process, let us consider a sequence $H_T = [(Q_1, A_1), ..., (Q_T, A_T)]$, where each $(Q_t, A_t)$ pair denotes a user query and the corresponding system response in the $t$-th round dialogue. At each step $t$, the system generates a response $A_t$ using the Loop Copilot function: 

$$A_t = \text{LoopCopilot}(Q_t, H_{t-1}).$$

Figure \ref{fig:diagram} shows the workflow of our proposed system. Loop Copilot comprises 5 key components: (1) The \textbf{large language model ($\mathcal{M}$)} for understanding and reasoning; (2) The \textbf{system principles ($\mathcal{P}$)} that provide basic rules to guide the large language model; (3) A list of \textbf{backend models ($\mathcal{F}$)} responsible for executing specific tasks; (4) A \textbf{global attribute table ($\mathcal{T}$)} that maintains crucial information to ensure continuity throughout the creative process; and (5) A \textbf{framework handler ($\mathcal{D}$)} that orchestrates the interactions between these components.

%

The workflow of Loop Copilot involves several steps. 

\begin{enumerate}
    \item \textbf{Input preprocessing.} the system processes the input by unifying the modality of the input. The framework handler $\mathcal{D}$ utilizes a music captioning model to describe the input music, while textual inputs are kept as they are.
    \item \textbf{Task analysis.} the framework handler performs task analysis if the text input contains an explicit demand. It calls the large language model $\mathcal{M}$ to analyze the task, resulting in a sequence of steps, which may involve a call to a single model or multiple chained calls to models, as the large language model may need to handle the task step by step. Section \ref{sec:supported_tasks} demonstrates the details.
    \item \textbf{Task execution.} After task analysis, the framework handler records all the steps and proceeds to execute the tasks. It calls the backend models in the specified order, providing them with the necessary parameters obtained from the large language model. If it requires a chained call of multiple models, the intermediate results generated by the previous model will be used in the next model.
    \item \textbf{Response generation. } Once the task execution is complete, the handler $\mathcal{D}$ collects the final result and sends it to the large language model for the final output. 
\end{enumerate}

Throughout this process, all operations are tracked and recorded in the global attribute table $\mathcal{T}$, ensuring consistency and continuity in the generation process. We will demonstrate it in detail in Section 3.3. Algorithm \ref{alg:workflow} illustrates the process during a T-round dialogue. 

\begin{algorithm}[htbp]
\caption{The workflow of Loop Copilot}\label{alg:workflow}
\renewcommand{\algorithmicrequire}{\textbf{Input:}}
\renewcommand{\algorithmicensure}{\textbf{Output:}}
\begin{algorithmic}
\Require user queries $Q = \{Q_1, ..., Q_T\}$
\Ensure responses $A = \{A_1, ..., A_T\}$
\State Initialize components: $\mathcal{M}, \mathcal{P}, \mathcal{F}, \mathcal{T}, \mathcal{D}$ 
\State Initialize chat history $H_0$ 
\State Define $A_0$ as initial music state or silence 
\For{$t$ in $[1, T]$} 
    \State $Q'_t \gets \mathcal{D}(Q_t)$ \Comment{\textbf{Input preprocessing}}
    \State $\mathcal{F}_{1:N} \gets \mathcal{M}(Q'_t, H_{t-1})$ \Comment{\textbf{Task analysis}}
    \State $A'_{t,0} \gets A_{t-1}$ \Comment{Initialize the chain}
    \For{$n$ in $[1, N]$} 
        \State $A'_{t, n} \gets \mathcal{F}_n(A'_{t,n-1})$ \Comment{\textbf{Task execution}}
    \EndFor
    \State $A_t \gets \mathcal{M}(A'_{t,N})$ \Comment{\textbf{Response generation}}
    \State $H_t \gets$ Append$(H_{t-1}, (Q_t, A_t))$  \Comment{Update chat history}
    \State Update $\mathcal{T}$ with key attributes from $A_t$ 
\EndFor
\end{algorithmic}
\end{algorithm}

\subsection{Supported Tasks} \label{sec:supported_tasks}

The interaction process within Loop Copilot is essentially a two-stage workflow, as illustrated in Figures \ref{fig:teaser} and \ref{fig:diagram}. The first stage involves the user drafting a music loop, while the second stage is dedicated to iterative refinement through dialogue. Each stage necessitates different tasks. In the initial stage, the focus is on creating music from an ambiguous demand, essentially a requirement for global features. The second stage shifts the focus to music editing, where fine-grained localized revisions are made. These revisions can include regenerating specific areas, adding or removing particular instruments, and incorporating sound effects. A comprehensive list of all supported tasks is presented in Table \ref{tab:tasks}.

Each task in Table \ref{tab:tasks} corresponds to one or more specific backend models, which are sequentially called as needed. For instance, consider the task ``impression to music". Here, a user can reference the title of a real-world music track. Loop Copilot first invokes ChatGPT to generate a description based on the given music title, which is then forwarded to MusicGen to generate the music audio. This ability to chain multiple models opens up a wealth of opportunities to accomplish new tasks that have scarcely been explored before, although the results may not be as good as for models trained for specific tasks. 

Specifically, we explore new methods for the tasks below:

\begin{enumerate}
    \item \textit{Imitate rhythmic pattern.} We utilise MusicGen's continuation feature to use the input drum pattern as a prefix while guiding the model with a target text description for generation.
    \item \textit{Impression to music.} For the `impression' descriptions that are not musical features but a reference to existing recordings, such as bands and titles, we first use ChatGPT to convert them into descriptions of musical features, and then call MusicGen to generate music audio. During the generation process, Loop Copilot does not directly copy music pieces from the original recording, therefore no IP problem will be raised.
    \item \textit{Add a track.} There are still no publicly available models supporting this feature. We instead utilise MusicGen's continuation feature to take the original audio as a prefix and use the new track text description to guide model generation. To ensure stability, we use the CLAP model to verify that the similarity between the generated result and the new text description is above a threshold.
\end{enumerate}

Note that Loop Copilot can comprehend complex demands that necessitate the combination of existing tasks. For instance, if a user wishes to ``generate jazz music and add medium level background noise, like in a pub", the large language model will dissect this demand into a series of tasks: ``text-to-music" and ``add sound effects". Within each task, if necessary, backend models are chained accordingly. Thus, the sequential invocation can occur at both the task and model levels. However, the final output presented to the user is the seamlessly integrated ``jazz music with background noise".

\begin{table*}[htbp]
\small
    \centering
    \begin{tabular}{lcll}
    \toprule
    Task & Stage & Examples of text input & Backend models  \\
    \midrule
    Text-to-music & 1 & Generate rock music with guitar and drums.  & MusicGen~\cite{musicgen} \\
    Drum pattern to music$^{\dag}$ & 1 & Generate rock music with guitar based on this drum pattern.  & MusicGen \\
    Impression to music$^{\dag}$ & 1 & Generate a music loop that feels like ``Hey Jude".  & ChatGPT, MusicGen \\
    Stylistic rearrangement & 1 & Rearrange this music to jazz with sax solo.  & MusicGen \\
    Music variation & 1 & Generate a music loop that sounds like this music.  & VampNet~\cite{vampnet} \\
    \midrule
    Add a track$^{\dag}$ & 2+ & Add a saxophone solo to this music loop. & MusicGen, CLAP~\cite{clap} \\
    Remove a track & 2+ & Remove the guitar from this music loop.  & Demucs~\cite{demucs} \\
    Re-generation/inpainting & 2+ & Re-generate the 3-5s part of the music loop. & VampNet \\
    Add sound effects & 2+ & Add some reverb to the guitar solo. & pedalboard~\tablefootnote{\url{https://doi.org/10.5281/zenodo.7817838}} \\ 
    Pitch shifting & 2+ & Transpose this music to G major. & pedalboard \\ 
    Tempo changing & 2+ & Make the music a bit slower. & torchaudio~\cite{torchaudio} \\ 
    \midrule
    Music captioning & N/A & Describe the current music loop.  & LP-MusicCaps \\
    \bottomrule
    \end{tabular}
    \caption{The list of all supported tasks in Loop Copilot at stage 1 (generation) and later stages (editing). We explore new training-free methods for those tasks with $\dag$ marks, as described in Section \ref{sec:supported_tasks}.}
    \label{tab:tasks}
\end{table*}

\subsection{Global Attribute Table}

The Global Attribute Table (GAT) is an integral component of the Loop Copilot system, designed to encapsulate and manage the dynamic state of music being generated and refined during the interaction process. Its role is to offer a centralized repository for the various attributes that define the musical piece at any given moment. This centralization is pivotal for Loop Copilot's ability to provide continuity, facilitate task execution, and maintain musical coherence. The design philosophy behind GAT draws inspiration from  ``blackboard" architectures~\cite{blackboard}. In this paradigm, the GAT can be likened to a blackboard---a shared workspace where different components of the system can access and contribute information. Table \ref{tab:gat} provides an example, showing the GAT state in the scenario of Figure \ref{fig:teaser}.

\begin{table}[htbp]
\small
    \centering
    \begin{tabular}{lp{2.5cm}ll}
    \toprule
     \textbf{bpm}         & 90 & \textbf{key}  & E$\flat$ major \\
     \midrule
     \textbf{genre}       & rock & \textbf{mood} & smooth \\
     \midrule
     \textbf{instruments} & \multicolumn{3}{l}{saxophone, guitar, snare drum} \\
     \midrule
     \textbf{description} & \multicolumn{3}{l}{\parbox{5.5cm}{smooth rock music loop with saxophone, a guitar arrangement and snare drum}} \\
     \midrule
     \multirow{2}{*}{\textbf{tracks}}   & \textbf{mix}  & \multicolumn{2}{l}{c540d5a6.wav} \\
     \cmidrule{2-4}
         & \textbf{stems} & \multicolumn{2}{l}{N/A} \\
    \bottomrule
    \end{tabular}
    \caption{An example of the Global Attribute Table in the scenario of Figure \ref{fig:teaser}.}
    \label{tab:gat}
\end{table}

GAT's significance can be further expounded upon through its multifaceted functionalities:

\begin{enumerate}
    \item \textbf{State Continuity:} GAT ensures that users experience a seamless dialogue with the Loop Copilot by persistently tracking musical attributes and evolving based on both user input and system output.
    
    \item \textbf{Task Execution:} During the task execution phase, backend models `$\mathcal{F}$` often require contextual information. GAT provides this context, thereby enhancing the models' performance.
    
    \item \textbf{Musical Coherence:} For any music creation tool, maintaining musical coherence is paramount. By storing key attributes like musical key and tempo, GAT ensures the harmonious and consistent evolution of music throughout the creative process.
\end{enumerate}

This collaborative approach ensures that all elements of Loop Copilot work in synergy, with GAT serving as the central point of reference, fostering an environment where every decision made is grounded in the broader context of the ongoing musical creation process.

\section{Experiments}

To evaluate the efficacy and usability of Loop Copilot, a mixed-methods experimental design is adopted, integrating both qualitative and quantitative research methods. This design aligns with the triangular research framework~\cite{triangle}. \footnote{Demo available at \url{https://sites.google.com/view/loop-copilot}.}

\subsection{Participants}

We recruited 8 volunteers (N=8) who were interested in AI-based music production, and work in the field of music and audio technology or production, though not necessarily professional-level musicians.
Participants provided informed consent, and data anonymization protocols were strictly followed to maintain ethical standards. The distribution of the participants was as follows:

\begin{enumerate}
    \item \textit{Experience in Music Production}: 3 starters (0-2 years), 3 intermediate (2-5 years), 2 experts (\textgreater 5 years).
    \item \textit{Experience in Music Performance}: 2 starters (0-2 years), 2 intermediate (2-5 years), 4 experts (\textgreater 5 years).
    \item \textit{Age}: 4 (18-35 years), 2 (35-45 years), 2 (\textgreater 45 years).
\end{enumerate}

\subsection{Measures}

We measure the following constructs: 

\begin{enumerate}
    \item \textit{Usability.} Usability serves as a critical metric for assessing the ease with which users can interact with Loop Copilot. It measures not only the system's efficiency but also gauges the intuitive nature of the user interface. We adopted the Standard System Usability Scale (SUS)~\cite{sus} (5-point Likert scale, see Appendix A) as a validated tool for this aspect of the evaluation. SUS scores have a range from 0 to 100, where a value over 68 is considered acceptable.
    \item \textit{Acceptance.} Understanding user acceptance is crucial for assessing whether Loop Copilot would be willingly incorporated into existing workflows. This encompasses factors like the perceived ease of use and the perceived usefulness of the system. The Technology Acceptance Model (TAM)~\cite{tam} served as the theoretical framework for evaluating these dimensions. Our TAM questionnaire (5-point Likert scale, see Appendix B) consists of 11 questions categorized into perceived ease of use  (Q1-4), perceived usefulness (Q5-8), and overall impressions (Q9-11).
    \item \textit{User experience.} Beyond usability and acceptance, the qualitative aspect of user experience provides a more nuanced understanding of the system's impact. This involves exploring the emotional and cognitive perceptions that users have when using Loop Copilot, such as the joys and frustrations they experience. Open-ended questions were designed to capture these subjective aspects in detail.
\end{enumerate}

\subsection{Procedure}

Experiments were conducted in a quiet, controlled environment to ensure consistency and minimize distractions. The experimental session for each participant consisted of three phases:

\begin{enumerate}
    \item \textit{Orientation Phase (10 minutes)}: 
    During this phase, participants were acquainted with the functionalities and features of Loop Copilot. This briefing aimed to standardize the initial level of understanding across participants.
    Specifically, the system was shown to the subjects with a brief explanation of how to use the interface.  Furthermore, the participants were presented with the example inputs in Table~\ref{tab:tasks} as examples of possible prompts supported by the system. 

    \item \textit{Interactive Usage Phase (20 minutes)}: 
    Participants were allowed to freely interact with Loop Copilot for music composition. Observational notes were made in real-time to capture immediate insights and identify areas for potential system improvement.

    \item \textit{Feedback and Evaluation Phase (15 minutes)}: 
    Upon completion of the interaction, participants were asked to fill out the Standard System Usability Scale (SUS) and Technology Acceptance Model (TAM) questionnaires. Additionally, a semi-structured interview based on the responses from the questionnaires was conducted to obtain qualitative feedback on their experience.

\end{enumerate}

Both quantitative (SUS, TAM scores) and qualitative (interview notes) data were collected. Data during the interview section were collected primarily through observational notes. These notes were aimed at capturing immediate insights, identifying potential areas for system improvement, and gathering qualitative feedback on the user experience. The choice of note-taking over audio recording was made to ensure participant anonymity and data privacy.

\subsection{Quantitive Results}

\subsubsection{System Usability Scale (SUS)}

\begin{figure}[htbp]
    \centering
    \includegraphics[trim={0.5cm 0 0.3cm 0}, width=\linewidth, clip]{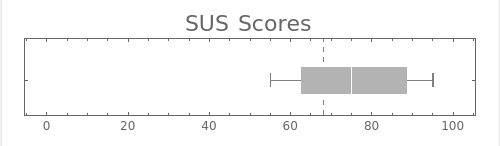}
    \caption{The box plot depicting SUS score results with an average of 75.31$\pm$15.32. The dotted line marks the threshold for effectiveness.}
    \label{fig:sus}
\end{figure}

The System Usability Scale (SUS) was used to measure the overall usability of Loop Copilot. The mean SUS score was \(75.31\) with a standard deviation of \(15.32\). According to the conventional SUS scale, a score above \(68\) is considered above average, suggesting that the participants found the system to be generally usable. A visualization is shown in Figure \ref{fig:sus}.

The SUS scores revealed a generally favorable perception of the system's usability. \footnote{For SUS scores, higher scores are better for odd-numbered problems and lower scores are better for even-numbered problems. The scores are converted and finally displayed as a score out of 100.} Users indicated a willingness to use the system frequently (Q1, 4.13$\pm$0.83), highlighting its perceived ease of use (Q3, 4.13$\pm$0.83) and quick learnability (Q7, 3.88$\pm$1.36).

However, some reservations were noted regarding the necessity for technical support (Q4, 2.63$\pm$1.41), suggesting that while the system is approachable, there may be layers of complexity that require expert guidance or better system onboarding; although the system’s features were generally considered well-integrated (Q5, 3.88$\pm$0.99), the middle-of-the-road scores for system consistency (Q6, 2.00$\pm$0.93) indicate room for improvement in unifying the system's functionalities. Such a sense of inconsistency may be correlated with the responsiveness of different AI models, which leads to the fact that some dialogues may have significantly longer response times than others.

\subsubsection{Technology Acceptance Model (TAM)}

\begin{enumerate}
    \item \textit{Perceived Usefulness (PU).} The average score for Perceived Usefulness was \(3.58\) with a standard deviation of \(1.13\). This indicates a moderate-to-high level of agreement among the participants that the system is useful. 
    \item \textit{Perceived Ease of Use (PEOU).} The average score for Perceived Ease of Use was \(3.89\) with a standard deviation of \(0.80\). This suggests that participants generally found the system easy to use.
    \item \textit{Overall TAM Scores.} The overall average TAM score was \(4.09\) with a standard deviation of \(1.09\), which suggests a favorable perception towards both the ease of use and usefulness of the system.
\end{enumerate}

\begin{figure}[htbp]
    \centering
    \includegraphics[trim={0 0 0.3cm 0}, width=\linewidth, clip]{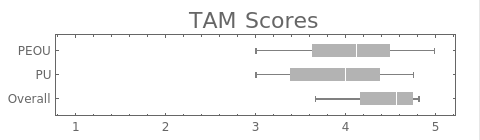}
    \caption{Box plot of the TAM score results. Perceived Usefulness (PU) with an average of 3.58$\pm$1.13; Perceived Ease of Use (PEOU) averaging 3.89$\pm$0.80, Overall TAM score of 4.09$\pm$1.09. These scores reflect participants' favorable perceptions of the system's utility and usability.}
    \label{fig:tam}
\end{figure}

A visualization is shown in Figure \ref{fig:tam}. 
TAM scores further solidified the system's positive impact on music performance, notably in terms of its usefulness (Q1-Q4, Q1: 4.25$\pm$0.89, Q2: 3.25$\pm$1.67, Q3: 4.13$\pm$0.64, Q4: 4.00$\pm$0.93) and user-friendly interface (Q5-Q8, Q5: 4.13$\pm$0.83, Q6: 4.63$\pm$0.52, Q7: 2.88$\pm$1.13, Q8: 4.63$\pm$0.74). The data indicated a strong inclination among users to integrate the system into their future workflows (Q9-Q11, Q9: 4.88$\pm$0.35, Q10: 4.75$\pm$0.46, Q11: 4.00$\pm$0.76), underscoring its perceived utility and ease of use.

\subsection{Qualitative Analysis}

Our qualitative analysis draws from an array of sources to form a nuanced view of the user experience. These include quantifiable metrics, user feedback, and observations gleaned during the interviews. We organize these insights into four broad categories: Overall Impressions, Positive Feedback, Areas of Concern, and Future Expectations.

\subsubsection{Overall Impressions} 

Participants generally found value in Loop Copilot as a tool for music generation. While the system was more favorably viewed for performance-oriented tasks rather than full-scale music production, users widely considered it a promising starting point for creative inspiration.
    
(1) Some participants found text-to-music conversion not fully meeting their specific musical visions, indicating a gap between user expectations and system output.

(2) Participants thought that Loop Copilot was useful for getting creative inspiration.
    
\subsubsection{Positive Feedback} 

\begin{enumerate}
    \item \textit{Ease of Use.} Most participants, especially beginners and intermediates, appreciated the intuitive nature of the interface. Most users found the system to be straightforward and easy to understand.

    \item \textit{Design and Interaction} Users lauded the design potential and interactive methods, suggesting that they represent a fertile ground for future development.
\end{enumerate}

\subsubsection{Areas of Concern}

\begin{enumerate}
    \item \textit{Limited Control and Precision.} Participants commonly mentioned the limited control they had over the musical attributes. Some cited specific instances where text prompts like \textit{``Add a rhythmic guitar"} or \textit{``Remove reverb"} were not adequately reflected in the output.
    \item \textit{Integration with Existing Workflows.} 
 Some users thought the system's current specifications were limited as a stand-alone music production system, and wanted it instead as a part of existing music creation systems, like a digital audio workstation.   
\end{enumerate}

\subsubsection{Future Expectations.}
\begin{enumerate}
    \item \textit{Feature Extensions.} Many users called for additional features like volume control, the ability to upload their own melody lines, and options for chord conditioning. Users also highlighted the need for multiple output options to choose from, rather than a single output.
    \item \textit{Improved Responsiveness.} Given that some participants found the system occasionally unresponsive to specific prompts, they hoped future versions could offer improved interpretation and execution of user commands.
\end{enumerate}

\section{Discussion}

\subsection{Limitations}

The quantitative results and the interviews suggest that our system is useful as an inspirational tool.  On the other hand, control of musical attributes can be improved, either by incorporating additional features into our system, or by allowing better coexistence with existing musical tools that allow fine-grained control.

In general, we found the freedom offered by an LLM to be a double-edged sword: it allows participants to explore freely to get musical inspiration, but without understanding the full capability of the system the user has little idea of how to get started.  Our experiments incorporated example prompts for onboarding the participants, which was essential for participants to get started in the interaction process.  This suggests LLM-based creation tools may benefit from providing hints on how to interact.

The user feedback illuminates several avenues for future work. First, enhancing user control over specific musical attributes could bridge the gap between user expectations and system output, such as chord conditioning. Second, integration with existing digital audio workstations was a frequent user request, suggesting that future versions could explore API-based integrations or even hardware-level compatibility. Users also expressed a desire for additional features like volume control and the ability to upload custom melody lines, as well as multiple output options for greater flexibility. Lastly, improved responsiveness to specific user prompts and the system's better tailoring for live performance versus production scenarios could also be areas for development.

\subsection{Potential Social Impact}

In developing Loop Copilot, we envision a platform that democratizes music creation, bridging gaps between expert musicians and enthusiasts. It can also foster greater diversity in music creation, as individuals from various backgrounds can now participate more actively without the traditional barriers of expensive equipment or years of training.

However, it is crucial to address the double-edged sword of AI-driven creative tools. On one hand, they can elevate amateur creations, but they may also inadvertently standardize musical outputs, potentially diluting the richness of human creativity. Furthermore, while our system promotes inclusivity, it is essential to ensure that it does not inadvertently reinforce cultural biases in music. For instance, the underlying models should be trained on diverse datasets to ensure a wide representation of global music genres.

Besides, with the potential integration of speech interactions, we anticipate enhancing accessibility, especially for users with visual or motor impairments. 


\section{Conclusion and future work}

In this paper, we presented Loop Copilot, a novel system that brings together Large Language Models and specialized AI music models to facilitate human-AI collaborative creation of music loops. Through a conversational interface, Loop Copilot allows for an interactive and iterative music creation process. We introduced a Global Attribute Table to keep track of the music's evolving state, ensuring that any modifications made are coherent and consistent. Additionally, we proposed a unique chaining mechanism that allows for training-free music editing by leveraging existing AI music models. Our evaluation, coupled with interview-based insights, demonstrates the potential of using conversational interfaces for iterative music editing.

As we look ahead, expanding Loop Copilot's functionalities stands out as a primary focus. Incorporating more intricate music editing tasks and specialized AI music models can cater to a broader range of musical preferences and genres. Additionally, transitioning to voice-based interactions offers advantage that enhancing accessibility for users with visual or motor impairments.



\section*{Acknowledgements}

We would like to acknowledge the use of free icons provided by Mavadee and Meaghan Hendricks in the diagram of this paper. Yixiao Zhang is a research student at the UKRI Centre for Doctoral Training in Artificial Intelligence and Music, supported jointly by the China Scholarship Council, Queen Mary University of London and Apple Inc.


\bibliographystyle{named}
\bibliography{ijcai24}

\appendix
\clearpage
\onecolumn
\section{SUS questionnaire}

\begin{enumerate}
    \item I think that I would like to use this system frequently.
    \item I found the system unnecessarily complex.
    \item I thought the system was easy to use.
    \item I think that I would need the support of a technical person to be able to use this system.
    \item I found the various functions in this system were well integrated.
    \item I thought there was too much inconsistency in this system.
    \item I would imagine that most people would learn to use this system very quickly.
    \item I found the system very cumbersome to use.
    \item I felt very confident using the system.
    \item I needed to learn a lot of things before I could get going with this system.
\end{enumerate}

\section{TAM questionnaire}

\begin{enumerate}
    \item I find Loop Copilot useful in live music performance.
    \item Using Loop Copilot improves my experience in music performance.
    \item Loop Copilot enables me to accomplish tasks more quickly.
    \item I find that Loop Copilot increases my productivity in music performance.
    \item I find Loop Copilot easy to use.
    \item Learning to operate Loop Copilot is easy for me.
    \item I find it easy to get Loop Copilot to do what I want it to do.
    \item I find the interface of Loop Copilot to be clear and understandable.
    \item Given the chance, I intend to use Loop Copilot.
    \item I predict that I would use Loop Copilot in the future.
    \item I plan to use Loop Copilot frequently.
\end{enumerate}

\section{ChatGPT Prompts}

\begin{center}
    \centering
    \begin{longtable}{p{0.24\linewidth}|p{0.75\linewidth}}
    \toprule
    Tool & Prompt \\ 
    \midrule\endfirsthead
    \toprule
    Tool & Prompt \\ 
    \midrule\endhead
      System prefix & Loop Copilot is designed to be able to assist with a wide range of text and music related tasks, from answering simple questions to providing in-depth explanations and discussions on a wide range of topics. Loop Copilot is able to generate human-like text based on the input it receives, allowing it to engage in natural-sounding conversations and provide responses that are coherent and relevant to the topic at hand.

    Loop Copilot is able to process and understand large amounts of text and music. As a language model, Loop Copilot can not directly read music, but it has a list of tools to finish different music tasks. Each music will have a file name formed as ``music/xxx.wav", and Loop Copilot can invoke different tools to indirectly understand music. When talking about music, Loop Copilot is very strict to the file name and will never fabricate nonexistent files. 
    
    Loop Copilot is able to use tools in a sequence, and is loyal to the tool observation outputs rather than faking the music content and music file name. It will remember to provide the file name from the last tool observation, if a new music is generated.
    
    Human may provide new music to Loop Copilot with a description. The description helps Loop Copilot to understand this music, but Loop Copilot should use tools to finish following tasks, rather than directly imagine from the description.
    
    Overall, Loop Copilot is a powerful music dialogue assistant tool that can help with a wide range of tasks and provide valuable insights and information on a wide range of topics.

    TOOLS:
    ------
    
    Loop Copilot has access to the following tools: \\
\midrule
    System format& To use a tool, you MUST use the following format:

\texttt{\textbf{Thought}: Do I need to use a tool? Yes}

\texttt{\textbf{Action}: the action to take, should be one of [\{tool\_names\}]}

\texttt{\textbf{Action Input}: the input to the action}

\texttt{\textbf{Observation}: the result of the action}

When you have a response to say to the Human, or if you do not need to use a tool, you MUST use the format:

\texttt{\textbf{Thought}: Do I need to use a tool? No}

\texttt{\{ai\_prefix\}\: [your response here]}\\
\midrule
System suffix & You are very strict to the filename correctness and will never fake a file name if it does not exist.
You will remember to provide the music file name loyally if it is provided in the last tool observation.

Begin!

Previous conversation history:
\{chat\_history\}

Since Loop Copilot is a text language model, Loop Copilot must use tools to observe music rather than imagination. The thoughts and observations are only visible for Loop Copilot.

\texttt{\textbf{New input}: \{input\}}

\texttt{\textbf{Thought}: Do I need to use a tool? \{agent\_scratchpad\}}

You MUST strictly follow the format. \\
\midrule
\midrule
    Text to music & 
    \textbf{Name}: Generate music from user input text.
    
    \textbf{Description}: useful if you want to generate music from a user input text and save it to a file. like: generate music of love pop song, or generate music with piano and violin.
    
    The input to this tool should be a string, representing the text used to generate music.\\
    \midrule
    Drum pattern to music & 
    \textbf{Name}: Generate music from user input text based on the drum audio file provided.

    \textbf{Description}: useful if you want to generate music from a user input text and a previous given drum audio file. like: generate a pop song based on the provided drum pattern above.
    
    The input to this tool should be a comma separated string of two, representing the music\_filename and the text description.
    \\
    \midrule
    Impression to music & 
    
    \textbf{Name}: Generate music from user input when the input is a title of music.

    \textbf{Description}: useful if you want to generate music which is silimar  and save it to a file. like: generate music of love pop song, or generate music with piano and violin.

    The input to this tool should be a comma separated string of two, representing the text description and the title.
    \\
    
    \midrule
    Stylistic rearrangement &  
    \textbf{Name}: Generate a new music arrangement with text indicating new style and previous music.

    \textbf{Description}: useful if you want to style transfer or rearrange music with a user input text describing the target style and the previous music. 
    
    Please use Text2MusicWithDrum instead if the condition is a single drum track. You shall not use it when no previous music file in the history. like: remix the given melody with text description, or doing style transfer as text described from previous music.
    
    The input to this tool should be a comma separated string of two, representing the music\_filename and the text description.\\
    \midrule
    Music variation & 
    \textbf{Name}: Generate a variation of given music.

    \textbf{Description}: useful if you want to generate a variation of music, or re-generate the entire music track. like: re-generate this music, or, generate a variant.
    
    The input to this tool should be a single string, representing the music\_filename.\\
    \midrule
    Add a track &  
    \textbf{Name}: Add a new track to the given music loop.

    \textbf{Description}: useful if you want to add a new track (usually add a new instrument) to the given music. like: add a saxophone to the given music, or add piano arrangement to the given music.
    
    The input to this tool should be a comma separated string of two, representing the music\_filename and the text description.\\
    \midrule
    Remove a track &  
    \textbf{Name}: Separate one track from a music file to extract (return the single track) or remove (return the mixture of the rest tracks) it.

    \textbf{Description}: useful if you want to separate a track (must be one of 'vocals', `drums', `bass', `guitar', `piano' or `other') from a music file. Like: separate vocals from a music file, or remove the drum track from a music file.
    
    The input to this tool should be a comma separated string of three params, representing the music\_filename, the specific track name, and the mode (must be `extract' or `remove').\\
    \midrule
    Re-generation/inpainting &  
    \textbf{Name}: Inpaint a specific time region of the given music.

    \textbf{Description}: useful if you want to inpaint or regenerate a specific region (must with explicit time start and ending) of music. like: re-generate the 3s-5s part of this music.
    
    The input to this tool should be a comma separated string of three, representing the music\_filename, the start time (in second), and the end time (in second).\\
    \midrule
    Add sound effects &  
    \textbf{Name}: Add a single sound effect to the given music. 

    \textbf{Description}: useful if you want to add a single sound effect, like reverb, high pass filter or chorus to the given music. like: add a reverb of recording studio to this music.
    
    The input to this tool should be a comma separated string of two, representing the music\_filename and the original user message.\\ 
    \midrule

    

    
    Pitch Shifting & 
    \textbf{Name}: Shift the pitch of the given music.

    \textbf{Description}: useful if you want to shift the pitch of a music. Like: shift the pitch of this music by 3 semitones.
    
    The input to this tool should be a comma separated string of two, representing the music\_filename and the pitch shift value.\\ 
    \midrule
    Speed Changing & 
    \textbf{Name}: Stretch the time of the given music.

    \textbf{Description}: useful if you want to stretch the time of a music. Like: stretch the time of this music by 1.5.
                    
    The input to this tool should be a comma separated string of two, representing the music\_filename and the time stretch value.\\ 
    \midrule
    Music captioning &  
    \textbf{Name}: Describe the current music.

    \textbf{Description}: useful if you want to describe a music. Like: describe the current music, or what is the current music sounds like.
    
    The input to this tool should be the music\_filename.\\ 

\bottomrule
\caption{List of system principles and task prompts. Each task features a unique name, description, and input parameter format for guiding the LLM.}
\label{tab:prompt}\\
    \end{longtable}

\end{center}
\clearpage
\twocolumn

\end{document}